\documentclass{aa}
\usepackage[varg]{txfonts}
\usepackage{graphicx,psfrag,amsmath,amssymb}
\usepackage[usenames,dvipsnames]{color}
\usepackage{epstopdf}
\usepackage{algorithm,algpseudocode}
\usepackage{fixltx2e}

\bibpunct{(}{)}{;}{a}{}{,} 
\def\bA{\mathbf{A}}

\def\bD{\mathbf{D}}
\def\bE{\mathbf{E}}

\def\bG{\mathbf{G}}
\def\bI{\mathbf{I}}

\def\bM{\mathbf{M}}
\def\bR{\mathbf{R}}
\def\bRh{\mathbf{\widehat{R}}}
\def\bW{\mathbf{W}}
\def\bX{\mathbf{X}}
\def\bZ{\mathbf{Z}}

\def\ba{\mathbf{a}}

\def\bd{\mathbf{d}}

\def\bg{\mathbf{g}}
\def\bn{\mathbf{n}}

\def\bs{\mathbf{s}}
\def\bw{\mathbf{w}}
\def\bx{\mathbf{x}}

\def\bz{\mathbf{z}}

\def\diag{\mathrm{diag}}
\def\argmin{\mathrm{argmin}}

\def\bDelta{{\mbox{\boldmath{$\Delta$}}}}
\def\bSigma{{\mbox{\boldmath{$\Sigma$}}}}

\def\bbeta{{\mbox{\boldmath{$\beta$}}}}

\def\bepsilon{{\mbox{\boldmath{$\epsilon$}}}}

\def\btheta{{\mbox{\boldmath{$\theta$}}}}
\def\bsigma{{\mbox{\boldmath{$\sigma$}}}}

\def\diag{\mathrm{diag}}

\def\expect{\mathcal{E}}
\def\Re{\mathrm{Re}}
\def\Im{\mathrm{Im}}
\def\trace{\mathrm{Tr}}
\def\bone{\mathbf{1}}
\def\bzero{\mathbf{0}}

\title{Fast gain calibration in radio astronomy using alternating direction implicit methods: Analysis and applications}
\titlerunning{Analysis of ADI methods for gain calibration}
\author{Stefano Salvini\inst{1} \and Stefan J.~Wijnholds\inst{2}}
\institute{Oxford e-Research Centre, 7 Keble Road,OX1 3QG, Oxford, United Kingdom, \email{stef.salvini@oerc.ox.ac.uk}
\and
Netherlands Institute for Radio Astronomy (ASTRON), P.O. Box 2, NL-7990 AA, Dwingeloo, The Netherlands, \email{wijnholds@astron.nl}}

\date{Received --, Accepted --}

\abstract{Modern radio astronomical arrays have (or will have) more than one order of magnitude more receivers than classical synthesis arrays, such as the VLA and the WSRT. This makes gain calibration a computationally demanding task. Several alternating direction implicit (ADI) approaches have therefore been proposed that reduce numerical complexity for this task from $\mathcal{O}(P^3)$ to $\mathcal{O}(P^2)$, where $P$ is the number of receive paths to be calibrated} 
{We present an ADI method, show that it converges to the optimal solution, and assess its numerical, computational and statistical performance. We also discuss its suitability for application in self-calibration and report on its successful application in LOFAR standard pipelines. } 
{Convergence is proved by rigorous mathematical analysis using a contraction mapping. Its numerical, algorithmic, and statistical performance, as well as its suitability for application in self-calibration, are assessed using simulations.} 
{Our simulations confirm the $\mathcal{O}(P^2)$ complexity and excellent numerical and computational properties of the algorithm. They also confirm that the algorithm performs at or close to the Cramer-Rao bound (CRB, lower bound on the variance of estimated parameters). We find that the algorithm is suitable for application in self-calibration and discuss how it can be included. We demonstrate an order-of-magnitude speed improvement in calibration over traditional methods on actual LOFAR data.} 
{In this paper, we demonstrate that ADI methods are a valid and computationally more efficient alternative to traditional gain calibration method and we report on its successful application in a number of actual data reduction pipelines.} 


\begin{document}
\maketitle

%
%
\section{Introduction}

Antenna-based gain calibration is a key step in the data reduction pipeline of any radio telescope. A commonly used method of estimating these antenna-based gains and possible other parameters in a (self-)calibration process is the Levenberg-Marquardt (LM) nonlinear least squares solver. Theoretically, the LM algorithm has at least $\mathcal{O}(N^3)$ complexity, where $N$ is the number of free parameters to be estimated. The LM solver has proved its value in self-calibration processes, but it is becoming a limiting factor in (near) real-time pipelines for modern telescopes, such as the Low Frequency Array (LOFAR, \citet{Vos2009ProcIEEE, Haarlem2013AandA}) and the Murchison Widefield Array (MWA, \citet{Lonsdale2009ProcIEEE, Bowman2013PASA}), owing to its cubic scaling with the number of receivers. The situation will only become worse for the Square Kilometre Array (SKA, \citet{Dewdney2009ProcIEEE, Dewdney2013SKA}).

This has motivated researchers to search for faster solvers with better scalability for antenna based gain calibration. \citet{Hamaker2000AandA} has already noted that solving for the gain of one specific receive path, assuming that all other receive paths are already calibrated while iterating over all antennas, could potentially be a fast way to solve for antenna-based gains in full polarization. This leads to an alternating direction implicit (ADI) method, which is used in the MWA real-time system for tile based calibration \citep{Mitchell2008JSTSP}. In the MWA pipeline, the gain estimates found for a given timeslice are used as initial estimates for the next timeslice. This makes a single iteration sufficient for achieving the required calibration accuracy. This cleverly exploits the electronic stability of the MWA system. \citet{Mitchell2008JSTSP} also proposed to reduce the noise on the estimates by using a weighted average between the current and previous gain estimates.

\citet{Salvini2011AAVP} showed that averaging the odd and even iterations not only reduces the noise on the estimates, but also considerably increases the rate of convergence and the robustness of the method. ADI methods  have $\mathcal{O}(P^2)$ complexity where $P$ is the number of receive paths to be calibrated. Since the number of visibilities also scales with $P^2$, these algorithms scale linearly with the number of data points and therefore have the lowest possible computational complexity for algorithms exploiting all available information.

Iterative algorithms, such as the ADI method presented in this paper, can be sensitive to the choice of initial estimates or exhibit slow convergence. In Sec.~\ref{sec:convergence} we therefore provide a rigorous convergence analysis. This gives a clear view of the algorithm's effectiveness and its potential limitations. We also discuss why these limitations are unlikely to hamper proper performance of the algorithm in practical radio astronomical applications, as by its actual use.

\citet{Hamaker2000AandA}, \citet{Mitchell2008JSTSP}, and \citet{Salvini2011AAVP} have derived the basic ADI iteration from the unweighted least squares cost function. In practice, weighted least squares methods are known to provide more accurate estimates if the S/N varies widely among data points. In this paper, we therefore start our derivation from the weighted least squares cost function and show that radio astronomical arrays can usually exploit the unweighted LS method.

In Sec.~\ref{sec:performance}, we compare the statistics of the gain estimates produced by the algorithm in Monte Carlo simulations with the Cramer-Rao bound (CRB). The CRB is the theoretical lower bound on the covariance of the estimated parameters obtained by an ideal unbiased estimator. The results indicate that the algorithm performs at the CRB when the covariance matched and unweighted least squares cost functions coincide (as expected) while performing very close to the CRB in most realistic scenarios. In the radio astronomical community, the ADI method presented in this paper is now usually referred to as StefCal, a name coined in jest by our colleagues Oleg Smirnov and Ilse van Bemmel. In view of its (close to) statistically efficient performance and high computational efficiency, we adapted the name to StEFCal, an admittedly rather contrived acronym for "statistically efficient and fast calibration".

StEFCal provides a considerable computational advantage over algorithms derived from the weighted least squares cost function, which usually scale with $P^3$ \citep{Ng1996TAP, Boonstra2003TSP, Wijnholds2009TSP}. In Sec.~\ref{sec:performance} we also consider the computational performance of StEFCal, highlighting its low computational complexity as well as its efficiency, its very small memory footprint, and scalability with
problem size.

In Sec.~\ref{sec:fullpol}, we briefly discuss the extension of StEFCal to full polarization calibration. This is now used routinely within MEqTrees \citep{Noordam2010AandA} and the LOFAR standard preprocessing pipeline \citep{Salvini2014CALIM}. Our simulations in Sec.~\ref{sec:performance} also show that StEFCal is suitable for integration in self-calibration approaches that rely on iterative refinement of the source model. In Sec.~\ref{sec:practice}, we show that StEFCal can be easily integrated in an actual system by reporting on the successful integration of StEFCal in a calibration pipeline for a LOFAR subsystem.

\begin{table}

\centering
\caption{Notation and frequently used symbols \label{tab:notation}}
\begin{tabular}{ll}
$a$ & scalar value\\
$\ba$ & vector $\ba$\\
$\bA$ & matrix $\bA$\\
$\bA_{:, k}$ & $k$-th column of the matrix $\bA$\\
$\diag \left ( \cdot \right )$ & converts a vector into a diagonal matrix\\
$\odot$ & Hadamard or element-wise product of matrices or\\
& vectors\\
$\left ( \cdot \right )^T$ & transpose\\
$\left ( \cdot \right )^H$ & Hermitian transpose\\
$\left ( \cdot \right )^{ [ i ]}$ & value  at the $i$-th iteration\\
$\expect \left \{ \bA \right \}$ & expected value of $\bA$\\ 
&\\
$\bRh$ & array covariance matrix, diagonal set to zero\\
$\bM$ & model cov.~matrix of observed scene, diagonal set\\
& to zero\\
$\bg$ & vector of complex valued receiver path gains\\
$\bG$ & $\bG = \diag \left ( \bg \right )$\\
$\bDelta$ & $\bDelta = \bRh - \bG \bM \bG^H$ \\
\end{tabular}

\end{table}

We conclude our paper in Sec.\ \ref{sec:discussion} by discussing possible alternative variants of the algorithm and possibilities for integrating StEFCal as a building block in other calibration algorithms, including algorithms dealing with direction-dependent gains, such as SAGECal \citep{Yatawatta2009DSP, Kazemi2011MNRAS} and the differential gains method proposed by \citet{Smirnov2011AandA}, corrupted or missing data values and polarization. For convenience of the reader, Table \ref{tab:notation} summarizes the notational conventions and frequently used symbols in this paper.

%
%
\section{Problem statement}

\subsection{Measurement equation}
\label{ssec:meq}

The radio astronomical system to be calibrated can have many different architectures. For example, antenna-based gain calibration can be applied to a synthesis array of dishes in interferometers, such as the VLA or the WSRT, but also to a synthesis array of stations in instruments, such as LOFAR or the envisaged Low Frequency Aperture Array (LFAA) system for the SKA \citep{Dewdney2013SKA}. Antenna-based gain calibration is also required within an aperture array station, where it becomes tile-based calibration in systems, such as the LOFAR high band antenna system \citep{Haarlem2013AandA} or the MWA. In this paper, we will therefore use generic terms, such as "receiving element", "element" or "antenna" to denote an individual element in a (synthesis) array instead of architecture-dependent terms such as "dish", "station" or "tile". We will also use the word "array" to refer to the system of elements to be calibrated instead of specific terms such as "station array", "synthesis array" or "tile array".

In this paper, we consider the scalar measurement equation or data model. The ADI method can be extended to full polarization as shown by \citet{Hamaker2000AandA}, \citet{Mitchell2008JSTSP} and \citet{Salvini2014CALIM, Salvini2014URSI}, but complicates the analysis unnecessarily. In our analysis, we assume that the source and noise signals are represented by complex valued samples that are mutually and temporally independent and that can be modeled as identically distributed Gaussian noise. We assume that these signals are spectrally filtered such that the narrowband condition \citep{Zatman1998ProcRSN} holds, which ensures that time delays can be represented by multiplication by phasors.

Besides allowing this representation of time delays, spectral filtering is a crucial step in (ultra-)wide band systems like modern radio telescopes for two other reasons. Firstly, it ensures that the noise in each channel can be assumed to be white noise regardless of bandpass fluctuations of the instrument or the inherent power spectrum of the observed sources. Secondly, observations using an increasingly larger fractional bandwidth are more likely to be affected by human-generated radio frequency interference (RFI). Most of this RFI can be detected and flagged \citep{Boonstra2005PhD, Offringa2012PhD, Offringa2013AandA}. If the channel width matches the bandwidth of the RFI signals (typically a few kHz), the S/N of the RFI in the occupied channel is maximized, thereby facilitating detection. As an additional bonus, the amount of spectrum that is flagged is minimized in this case. RFI that escapes detection can cause outliers in the measured data that do not fit a Gaussian noise model. In such cases, an appropriate weighting of the data samples can help to improve robustness to such outliers \citep{Kazemi2013MNRAS}. In Sec.~\ref{sec:discussion} we briefly discuss how such weighting can be incorporated in StEFCal, although at the expense of some computational efficiency.

The direction-independent gain of the $p$-th receive path of an array consisting of $P$ elements can be represented by the complex valued scalar $g_p$. The output signal of the $p$-th element, receiving signals from $Q$ sources, as a function of time $t$ can be described by
\begin{equation}
x_p(t) = g_p \sum_{q=1}^Q a_{p,q} s_q ( t ) + n_p (t), \label{eq:sensor_response}
\end{equation}
where $a_{p,q}$ is the $p$-th antenna's response to the $q$-th source, $s_q ( t )$ is the source signal, and $n_p ( t )$ represents the noise on the measurement. Since we assume that the narrowband condition holds, the factors $a_{p,q}$ are the geometry-dependent phase terms that result in the familiar Fourier-transform relationship between the source structure and the measured visibilities \citep{Thompson2004}.

We can stack the element-indexed quantities in vectors of length $P$:
the output signals of the individual sensors in
$\bx ( t ) = \left [ x_1  ( t ), \cdots, x_P ( t ) \right ]^T$; the complex valued gains in $\bg = \left [ g_1, \cdots, g_P \right ]^T$;
the array response vector to the $q$-th source as $\ba_q =\left [ a_{1, q}, \cdots, a_{P, q} \right ]^T$;
finally, the noise vector as \mbox{$\bn ( t ) = \left [ n_1 (t), \cdots, n_P ( t ) \right ]^T$}.

With these definitions, we can describe the array signal vector as
\begin{equation}
\bx (t) = \bg \odot \sum_{q=1}^Q \ba_q s_q (t) + \bn ( t ). \label{eq:array_sig_vec}
\end{equation}
Defining the $P \times P$ diagonal matrix $\bG = \diag (\bg)$, the  $P \times Q$ array response matrix $\bA  = \left [ \ba_1, \cdots, \ba_Q \right ]$, and the $Q \times 1$ source signal vector $\bs (t) = \left [ s_q (t), \cdots, s_Q  (t) \right ]^T$, we can reformulate \eqref{eq:array_sig_vec} in a convenient matrix form:
\begin{equation}
\bx (t) = \bG \bA \bs (t) + \bn (t). \label{eq:array_sig_vec2}
\end{equation}

Defining $\bX = \left [ \bx (T), \cdots, \bx ( K T) \right ]$, where $K$ is the number of samples and $K T$ defines the overall measurement duration, we can then estimate the array covariance matrix, often referred to as visibility matrix or matrix of visibilities, by
\begin{equation}
\widehat{\bR} = \frac{1}{K} \bX \bX^H.
\end{equation}
The model for the array covariance matrix follows from
\begin{equation}
\bR = \expect \left \{ \frac{1}{K} \bX \bX^H \right \} = \bG \bA \bSigma_s \bA^H \bG^H + \bSigma_n, \label{eq:acm}
\end{equation}
where $\bSigma_s = \expect \left \{ \bs (t) \bs^H (t) \right \}$ is the  covariance matrix of the source signals, and $\bSigma_n = \expect \left \{ \bn \left ( t \right  ) \bn^H (t) \right \}$ is the noise covariance matrix. In \eqref{eq:acm}, we have assumed that the source and noise signals are mutually uncorrelated. We also assume that the noise signals on the individual sensors are uncorrelated, such that the noise covariance matrix is diagonal, i.e., $\bSigma_n = \diag ( \bsigma_n )$. In Sec.\ \ref{sec:discussion}, we indicate how the algorithm can deal with more complicated noise models.
For convenience of notation, we introduce \mbox{$\bR_0 = \bA \bSigma_s \bA^H$}, so that we can write \eqref{eq:acm} as
\begin{equation}
\bR = \bG \bR_0 \bG^H + \bSigma_n.
\end{equation}

%
%
\subsection{Optimization problem}
\label{ssec:optim}

The antenna based gains and phases can be calibrated by a measurement in which the source structure is known, so we can predict model visibilities $\bR_0$. Since the receiver path noise powers $\bsigma_n$ are usually not known, the calibration problem is described by
\begin{equation}
\left \{ \widehat{\bg}, \widehat{\bsigma}_n  \right \} = \underset{\bg,\bsigma_n}{\argmin} \left \| \bW^H \left ( \widehat{\bR} - \bG \bR_0 \bG^H - \bSigma_n \right ) \bW \right \|_F^2. \label{eq:problem_statement}
\end{equation}
This equation describes our problem as a weighted least squares estimation problem. This allows us to apply covariance matched weighting by taking $\bW = \bR^{-1/2}$, leading to estimates that are asymptotically, for a large number of samples, equivalent to maximum likelihood estimates \citep{Ottersten1998DSP}. However, in radio astronomy, sources are typically much weaker than the noise, i.e., the S/N per sample is usually very low. Exceptions to this statement are observations of the brightest sources on the sky, such as Cas A, Cyg A, and the Sun, in which self-noise becomes a significant issue \citep{Kulkarni1989AJ, Wijnholds2010PhD}. Besides such exceptional cases, the model covariance matrix can be approximated by $\bR \approx \bSigma_n$. Since many radio astronomical instruments are arrays of identical elements, whereby $\bSigma_n \approx \sigma_n \bI$, we are justified in using $\bW = \bI$. In the Monte Carlo simulations presented in Sec.~\ref{sec:sims}, we demonstrate that violating these assumptions only leads to small deviations from the CRB, even in extreme situations unlikely to occur in reality.

In many practical cases, we have an incomplete model of the observed field, and we employ the best available model $\bM \approx \bR_0$ which, for example, only includes the brightest sources. In our simulations, we consider both complete and incomplete information on the observed field.
Another practical matter is that the autocorrelations are dominated by the noise power of the array elements. Since accurate modeling of the diagonal of the array covariance matrix involves estimating the noise power of each individual element, we are forced to estimate the antenna-based gains using the crosscorrelations followed by estimation of the noise powers using the diagonal elements. For the gain estimation step, it is therefore convenient to set the diagonal entries of $\widehat{\bR}$ and $\bM$ to zero. This assumption is made throughout this paper, thus ignoring $\bSigma_n$. This simplifies the estimation problem described in \eqref{eq:problem_statement} to
\begin{equation}
\widehat{\bg} = \underset{\bg}{\argmin} \left \| \widehat{\bR} - \bG \bM \bG^H \right \|_F^2 = \underset{\bg}{\argmin} \left \| \bDelta \right \|_F^2
\end{equation}
where we have introduced $\bDelta = \bRh - \bG \bM \bG^H$ for brevity of notation.

%
%
\section{The algorithm}
\label{sec:algorithm}

Using an alternating direction implicit (ADI) approach, we first solve for $\bG^H$ holding $\bG$ constant, then for $\bG$ holding $\bG^H$ constant. Since $\bDelta$ is Hermitian, the two steps are equivalent and the iteration consists of only the following step:
\begin{equation}
\bG^{[i]} = \underset{\bG}{\argmin} ~ {\|  \bRh - {\bG}^{[i-1]} \bM \bG^H \|}_F \; .
\end{equation}
Since $\| \bx \|_F = \| \bx \|_2$ for any vector $\bx$, and setting
\begin{equation}
\bZ^{[i]} = \bG^{[i]} \bM
\label{eq:adi:basic}
\end{equation}
we can write
\begin{equation}
{\left \|  \bDelta  \right \|}_F = {\left \| \bRh - \bZ \bG^H \right \|}_F = \sqrt {\sum_{p = 1}^{P} {\left \| {\bRh_{:, p} - \bZ_{:, p} g_p^*} \right \|}_2^2} \;.
\label{eq:adi:basic2}
\end{equation}

Equation \eqref{eq:adi:basic2} shows that the complex gains $g_p$ are decoupled and that each iteration consists of solving $P$ independent $P \times 1$ linear least squares problems. Using, for example, the normal equation method we readily obtain:
\begin{equation}
g_p^{[ i ]} = \left \{ \frac{(\bZ^{[ i-1 ]}_{:, p} )^H \cdot \bRh_{:, p}} 
{(\bZ^{[ i-1 ]}_{:, p} )^H \cdot \bZ^{[ i-1 ]}_{:, p} } \right \}^* = \frac{\bRh_{:, p}^H \cdot \bZ^{[i-1]}_{:, p}}{(\bZ^{[ i-1 ]}_{:, p} )^H \cdot \bZ^{[ i-1 ]}_{:, p} } \; ,\label{eq:adi:basic3}
\end{equation} 
which is the basic ADI iteration.

In practice, the basic iteration may converge very slowly. For example, in the case of the sky model used in Sec.~\ref{sec:performance}, it does not converge at all, bouncing to and fro between two vectors $\bg$. We resolved this issue by replacing the gain solution of each {\em even} iteration by the average of the current gain solution and the gain solution of the previous {\em odd} iteration. This simple process makes the iteration both very fast and very robust. This is the basic variant of the StEFCal algorithm, which is described here as Algorithm \ref{alg:stefcal}. Its convergence properties are studied in detail in the next section, while its numerical, computational, and statistical performance are discussed in Sec.\ \ref{sec:performance}.

\begin{algorithm}
\caption{Algorithm StEFCal}
\begin{algorithmic}
\State Initiate $\bG^{[0]}$;  $\bG^{[0]} = \bI$ is adequate in most cases
\For { $i  =  1, 2, \cdots , i_\mathrm{max}$} 
\For { $p  =  1, 2, \cdots , P $} 
\State $\bz \gets \bG^{[i-1]}  \cdot \bM_{:, p}  \equiv \bg^{[i-1]} \odot \bM_{ :, p }$
\State $g_p \gets (\bRh_{:, p}^H \cdot \bz ) / (\bz^H \cdot \bz)$
 \EndFor
 \If {$ \mathrm{mod}_{2} (i) = 0$}
\If {$\| \bg^{[i]} - \bg^{[i-1]} \|_F / \| \bg^{[i]} \|_F \leq \tau$} 
\State Convergence reached
\Else
\State $\bG^{[i]} \gets ( \bG^{[i]} + \bG^{[i-1]} ) / 2$
\EndIf
\EndIf
\EndFor
\end {algorithmic}
\label{alg:stefcal}
\end {algorithm}

We want to stress a few important points
\begin{itemize}
\item There are no parallel dependencies in the inner loop (the dependency on $\bz$ is trivially resolved by employing a local vector $\bz$ on each computational unit or core, or by using the individual elements of $\bz$ at once without needing to store the vector, although at the cost of potentially lower performance).
\item All data are accessed through unit strides (contiguous memory locations).
\item The memory footprint is very small.  Basically, only one extra $P$-vector is required (for requirements of $\bz$ see above) besides the visibility matrices.
\end{itemize}
Thus, the algorithm can be readily implemented on many-core architectures (such as GPUs, Intel Xeon Phi, etc.), as well as on multiple cores, for example by employing OpenMP. Codes and algorithms are in a very advanced development stage and are available on request.

The gain estimation problem has an inherent phase ambiguity. In this paper, we choose, entirely arbitrarily, to use the first receiver as phase reference. This constraint can be imposed either within each iteration at the cost of $\mathcal{O}(P)$ operations or at the end of the computation. In practical terms, we did not find any difference in rate of convergence and results between these two options.

We now look at the algorithm in terms of the gradient with respect of the real and imaginary parts of the complex gains of the function
\begin{equation}
\widehat{\bg} = \underset{\bg}{\mathrm{argmin}} \left \| \bDelta \right \|_F^2  ~=~ \underset{\bg}{\mathrm{argmin}} ~\trace \left ( \bDelta \bDelta^H \right ).
\end{equation}

At a minimum, the partial derivatives with respect to the real and imaginary parts of the complex gains must all be zero. Hence,
\begin{flalign}
& \lefteqn{\frac{\partial}{\partial~\Re (g_p)} \trace \left ( \bDelta \bDelta^H \right ) =
     \trace \left [ \frac{\partial~\bDelta \bDelta^H}{\partial~ \Re (g_p)} \right ] =} & \nonumber\\
& \; = \; - \trace \left [ \left ( \bE_p  \bM \bG^H + \bG \bM \bE_p  \right ) \bDelta^H + \mathrm{c.c.} \right ] & \nonumber\\
& \; = \; -2 \Re \left [ \trace \left ( \bE_p \bM \bG^H \bDelta^H \right ) + \trace \left ( \bG \bM \bE_p \bDelta^H \right )\right ] & \nonumber\\
& \; = \; -4 \Re \left \{ \trace \left ( \bDelta \bE_p \bZ^H \right ) \right \} \nonumber\\
& \; = \; -4 \Re \left [ \bZ^H_{:,p} \bDelta_{:, p} \right ] & \nonumber \\
& \; = \; -4 \Re \left [ \bZ^H_{:, p} \cdot \left ( \bRh_{:, p} - \bZ_{:, p} g_p^* \right ) \right ] & \nonumber\\
& \; = \; 0 &
\label{eq:partial_cal_real}
\end{flalign}
where $\mathrm{c.c.}$ stands for complex conjugate and $\bE_p$ denotes the $P \times P$ elementary matrix, which only contains zeros except for the $(p, p)$-element, which is unity. We used the properties of the trace of a product of matrices; we used the Hermitian properties of $\bDelta$, $\bM$, and $\bE_p$; and finally, we used $\bZ_{:, p} = \left (\bG \bM  \right )_{:, p}$. Likewise, for the imaginary part of $g_p$ we obtain
\begin{equation}
\frac{\partial \trace \left ( \bDelta \bDelta^H \right )}{\partial~\Im (g_p)} =  -4 \Im \left [ \bZ^H_{:, p} \cdot \left ( \bRh_{:, p} - \bZ_{:, p} g_p^* \right ) \right ] = 0
\; .
\label{eq:partial_cal_imag}
\end{equation}

Looking at \eqref{eq:adi:basic2}, \eqref{eq:partial_cal_real}, and \eqref{eq:partial_cal_imag}, and at Algorithm \ref{alg:stefcal}, we can see that the termination condition in the algorithm implies zero gradient as a function of the real and imaginary parts of all $g_p$, for $p = 1, \cdots, P$, achieved through a process of local minimization via a linear least squares method.  Because the algorithm shows very good convergence in all realistic cases studied, we can infer that StEFCal does indeed produce gains that minimize $\bDelta$ in the least squares sense.

Moreover, using Eqs.~\eqref{eq:partial_cal_real} and \eqref{eq:partial_cal_imag}, we can obtain the components of the gradient with respect to the real and imaginary part of $\bg$ at the $i$-th iteration by
\begin{eqnarray}
\nabla_{\Re (g_p)} \left \| \bDelta \right \|^2_F & = & -4 \Re \left [ \bz_p^{[i-1] H} \bRh_{:, p} - \left ( \bz_p^{[i-1] H} \bz_p^{[i-1]} \right) g_p^{[i-1]*} \right ] \nonumber \\
\nabla_{\Im (g_p)} \left \| \bDelta \right \|^2_F & = & -4 \Im \left [ \bz_p^{[i-1] H} \bRh_{:, p} - \left ( \bz_p^{[i-1] H}  \bz_p^{[i-1]} \right) g_p^{[i-1]*}  \right ]\nonumber
\end{eqnarray}
where we used the notation $\bz_p^{[i-1]} = \bZ_{:,p}^{[i-1]} = \bg^{[i-1]} \odot \bM_{ :, p }$. The dot products have already been computed to generate the new gain estimate $g_p^{[i]}$, so the components of the gradient can be generated at virtually no cost.

%
%
\section{Analysis of convergence}
\label{sec:convergence}

In this section we first introduce the concept of contraction mapping and then employ this concept to analyze the convergence properties of the proposed algorithm. The special case of calibration on a single point source shows that the algorithm converges for all initial estimates except for initial estimates in the null space of $\bg$. Finally, we study the general case of an arbitrary source distribution, showing that convergence is achieved when certain conditions on the observed scene and the initial estimate are met. We discuss these conditions and argue that they are met in practical situations. The convergence analysis presented below considers convergence in the noise-free case. The effect of measurement noise is studied in detail using Monte Carlo simulations in Sec.~\ref{sec:sims}.

%
%
\subsection{Condition for convergence}

A contraction mapping on a complete metric space $\mathcal{M}$ with distance measure $d$ is a function
$f:\mathcal{M} \rightarrow \mathcal{M}$ with the property that a real valued number $\mu < 1$ exists such that for all $x, y \in \mathcal{M}$
\begin{equation}
d \left ( f \left ( x \right ), f \left ( y \right ) \right ) \leq \mu d\left (x, y \right ).
\end{equation}
The Banach fixed point theorem states that the sequence of values resulting from iterative application of a contraction mapping converges to a fixed point \citep{Palais2007JFPTA}. This theorem can be understood intuitively. We consider two arbitrary points in a Euclidian space using the induced norm of their difference to measure the distance between them. If a given operation is a contraction mapping, applying that operation to both points separately will produce two new points that are closer together. Repeated application of the operation on the resulting points will make the distance between each pair of new points shorter than the previous pair. If we continue applying the operation long enough, we can make this distance arbitrarily small, thus effectively converging to a single point. If we can show that the full iteration (two basic iteration as described by \eqref{eq:adi:basic3} and the averaging step) is a contraction mapping, we can conclude that the iterative application of the full iteration leads to a converging sequence of values $\bg^{[i]}$.

Replacing $\bRh$ with $\bR = \bG \bM \bG^H = \bg \bg^H \odot \bM$, the basic iteration for a single element described by \eqref{eq:adi:basic3} reads as
\begin{equation}
g_p^{[i]} = \frac{\left ( \bg \odot \bM_{:,p} \right )^H \left ( \bg^{[i-1]} \odot \bM_{:,p} \right )}{\left ( \bg^{[i-1]} \odot \bM_{:,p} \right )^H \left ( \bg^{[i-1]} \odot \bM_{:,p} \right )} g_p.
\end{equation}
Introducing the weight vector $\bw_p = \bM^*_{:,p} \odot \bM_{:,p}$, we can read the products in the numerator and denominator as weighted inner products and write the basic iteration as
\begin{equation}
g_p^{[i]} = \frac{\left \langle \bg^{[i-1]}, \bg \right \rangle_{\bw_p}}{\left \langle \bg^{[i-1]}, \bg^{[i-1]} \right \rangle_{\bw_p}} g_p \label{eq:basic_iter_inner_prod}.
\end{equation}

The initial estimate $\bg^{[i-1]}$ can be written in terms of a scaling $\alpha$ of the true gain values $\bg$ and an error vector orthogonal to $\bg$ (in the usual Euclidean sense) $\bepsilon$; i.e., we may write $\bg^{[i-1]} = \alpha \left ( \bg + \bepsilon \right )$. Substitution in \eqref{eq:basic_iter_inner_prod} gives
\begin{equation}
g_p^{[i]} = \frac{1}{\overline{\alpha}} \frac{ \left \langle \bg + \bepsilon, \bg \right \rangle_{\bw_p}}{\left \Arrowvert \bg + \bepsilon \right \Arrowvert^2_{\bw_p}} g_p =\frac{1}{\overline{\alpha}} \beta_p g_p. \label{eq:basic_iter_inner_prod2}
\end{equation}
This formulation gives interesting insight into the operation of the basic iteration. If the initial estimate is purely a scaling of the true value, $\beta_p = 1$, and the algorithm only tries to adjust the amplitude. If the initial estimate has a component that is orthogonal to the true gain vector, the algorithm tries to remove $\bepsilon$ by projecting the guessed gain vector on the true gain vector. The impact of $\bepsilon$ depends on the element being considered, because the scene used for calibration may be such that the calculation of $\beta_p$ involves geometry in a weighted Euclidean space.

Introducing the vector $\bbeta = \left [ \beta_1, \cdots, \beta_P \right ]^T$, we can write the full iteration for a single element as
\begin{equation}
g_p^{[i]} = \frac{1}{2} \left ( \frac{\left \langle \frac{1}{\overline{\alpha}} \bbeta \odot \bg, \bg \right \rangle_{\bw_p}}{\left \Arrowvert \frac{1}{\overline{\alpha}} \bbeta \odot \bg \right \Arrowvert^2_{\bw_p}} g_p + \frac{1}{\overline{\alpha}} \beta_p g_p \right ).
\end{equation}
For the gain vector, the full iteration is described by
\begin{equation}
\bg^{[i+1]} = \frac{1}{2} \left ( \alpha \widetilde{\bbeta} \odot \bg + \frac{1}{\overline{\alpha}} \bbeta \odot \bg \right ) = \frac{1}{2} \left ( \frac{\left | \alpha \right | \widetilde{\bbeta} + \bbeta}{\overline{\alpha}} \right ) \odot \bg,
\end{equation}
where we defined $\widetilde{\beta}_p = \left \langle \bbeta \odot \bg, \bg \right  \rangle_{\bw_p} / \left \Arrowvert \bbeta \odot \bg \right \Arrowvert^2_{\bw_p}$ (note the similarity in form and hence interpretation as $\beta_p$) and the vector $\widetilde{\bbeta} = \left [ \widetilde{\beta}_1, \cdots, \widetilde{\beta}_P \right ]^T$ for brevity of notation. Although not recognizable in this equation, the initial estimate $\bg^{[i-1]}$ comes in via $\alpha$, $\bbeta$ and $\widetilde{\bbeta}$.

For convergence, we like to show that, if we have two distinct initial estimates $\bg_1^{[0]}$ and $\bg_2^{[0]}$, we have
\begin{equation}
\left \Arrowvert \bg_1^{[2]} - \bg_2^{[2]} \right \Arrowvert \leq \left \Arrowvert \bg_1^{[0]} - \bg_2^{[0]} \right \Arrowvert, \label{eq:convergence_req}
\end{equation}
where we used the Euclidian norm (induced norm) as distance measure in the linear space of potential gain vectors. Since we are considering the change in Euclidian distance between these two initial estimates, we can attribute the differential error vector to one of the gains without loss of generality, and model the initial estimates as $\bg_1^{[0]} = \alpha_1 \left ( \bg + \bepsilon \right )$ and $\bg_2^{[0]} = \alpha_2 \bg$, leading to
\begin{equation}
\left \Arrowvert \frac{1}{2} \left ( \frac{\left | \alpha_1 \right | \widetilde{\bbeta}_1 + \bbeta_1}{\overline{\alpha_1}} - \frac{\left | \alpha_2 \right | \widetilde{\bbeta}_2 + \bbeta_2}{\overline{\alpha_2}} \right ) \odot \bg \right \Arrowvert \leq \left \Arrowvert \alpha_1 \left ( \bg + \bepsilon \right ) - \alpha_2 \bg \right \Arrowvert. \label{eq:convergence_req1}
\end{equation}
Since $\bepsilon = 0$ for $\bg_2^{[0]}$, we have $\bbeta_2 = \widetilde{\bbeta}_2 = \bone$. This allows us to simplify the lefthand side of \eqref{eq:convergence_req1} to
\begin{eqnarray}
\lefteqn{\left \Arrowvert \frac{1}{2} \left ( \frac{\overline{\alpha}_2 \left | \alpha_1 \right | \widetilde{\bbeta}_1 + \overline{\alpha}_2 \bbeta_1 - \overline{\alpha}_1 \left | \alpha_2 \right | - \overline{\alpha}_1}{\overline{\alpha}_1 \overline{\alpha}_2} \right ) \odot \bg \right \Arrowvert} \nonumber\\
& = & \frac{1}{2} \left \Arrowvert \diag \left ( \alpha_1 \widetilde{\bbeta}_1 - \alpha_2 + \frac{\overline{\alpha}_2 \bbeta_1 - \overline{\alpha}_1}{\overline{\alpha}_1 \overline{\alpha}_2} \right ) \bg \right \Arrowvert \nonumber\\
& \leq & \frac{1}{2} \sigma_{\mathrm{max}} \left ( \diag \left ( \alpha_1 \widetilde{\bbeta}_1 - \alpha_2 + \frac{\overline{\alpha}_2 \bbeta_1 - \overline{\alpha}_1}{\overline{\alpha}_1 \overline{\alpha}_2} \right ) \right ) \left \Arrowvert \bg \right \Arrowvert,
\label{eq:convergence_req1a}
\end{eqnarray}
where $\sigma_{\mathrm{max}}(\cdot)$ denotes the largest singular value of a matrix. For a diagonal matrix $\bD$, such as that in \eqref{eq:convergence_req1a}, we have:
\begin{equation}
\sigma_{\mathrm{max}}^2 \left ( \bD \right ) = \lambda_{\mathrm{max}} \left ( \bD^H \bD \right ) = \mathrm{max}_n \left \{ \left | \bd_n \right |^2 \right \},
\end{equation}
where $\bd_n$ denotes the $n$th element on the main diagonal of $\bD$.

Squaring the left- and righthand sides of \eqref{eq:convergence_req} and exploiting the fact that $\bepsilon \perp \bg$, we require
\begin{eqnarray}
\lefteqn{\frac{1}{4} \left | \alpha_1 \widetilde{\beta}_{1,p_\mathrm{max}} - \alpha_2 + \frac{\overline{\alpha}_2 \beta_{1,p_\mathrm{max}} - \overline{\alpha}_1}{\overline{\alpha}_1 \overline{\alpha}_2} \right |^2} \nonumber\\
&&\leq \left | \alpha_1 - \alpha_2 \right |^2 + \left | \alpha_1 \right |^2 \frac{\left \Arrowvert \bepsilon \right \Arrowvert^2}{\left \Arrowvert \bg \right \Arrowvert^2} \label{eq:convergence_req2}
\end{eqnarray}
for convergence, where $p_\mathrm{max}$ is the value of $p$ that maximizes the lefthand side.

%
%
\subsection{Convergence for single point source calibration}

When the observed scene consists of a single point source, we can assume $\bM = \bR_0 = \bone \bone^H$ without loss of generality. The weighted inner products in $\bbeta$ and $\widetilde{\bbeta}$ therefore reduce to the standard Euclidean inner product since $\bw_p = \bone$ for all $p$. Since the error vector $\bepsilon$ is assumed to be perpendicular to $\bg$, \eqref{eq:basic_iter_inner_prod2} shows that $\bepsilon$ will be projected out in the first basic iteration. As a result, we have $\beta_p = 1$ for all $p$ after the first iteration, which reduces the condition for convergence in \eqref{eq:convergence_req2} to
\begin{equation}
\frac{1}{4} \left | \alpha_1 - \alpha_2 + \frac{\overline{\alpha}_2 - \overline{\alpha}_1}{\overline{\alpha}_1 \overline{\alpha}_2} \right |^2 \leq \left | \alpha_1 - \alpha_2 \right |^2 . \label{eq:convergence_req_pointsrc}
\end{equation}
Substitution of
\begin{eqnarray}
\lefteqn{\left | \alpha_1 - \alpha_2 + \frac{\overline{\alpha}_2 - \overline{\alpha}_1}{\overline{\alpha}_1 \overline{\alpha}_2} \right |^2}\nonumber\\
& \leq & \left ( \left | \alpha_1 - \alpha_2 \right | + \frac{\left | \overline{\alpha}_2 - \overline{\alpha}_1 \right |}{\left | \overline{\alpha}_1 \right | \left | \overline{\alpha}_2 \right |} \right )^2 \nonumber\\
& = & \left | \alpha_1 - \alpha_2 \right |^2 + \frac{\left ( 2 \left | \alpha_1 \right | \left | \alpha_2 \right | + 1 \right ) \left | \alpha_1 - \alpha_2 \right |^2}{\left | \alpha_1 \right |^2 \left | \alpha_2 \right |^2}
\end{eqnarray}
in \eqref{eq:convergence_req_pointsrc}, followed by some algebraic manipulation, gives
\begin{equation}
2 \left | \alpha_1 \right | \left | \alpha_2 \right | + 1 \leq 3 \left | \alpha_1 \right |^2 \left | \alpha_2 \right |^2, \label{eq:convergence_scaling}
\end{equation}
which holds if $\left | \alpha_1 \right | \left | \alpha_2 \right |\geq 1$. This may not be true for the initial estimate provided to the algorithm, but as long as the initial estimate does not lie in the null space of $\bg$, this condition will be met after the first full iteration, since $\frac{1}{2} \left | 1 / \overline{\alpha} + \alpha \right |\geq 1$ for all values $\alpha$ satisfying $\left | \alpha \right |\geq 0$. Equation \eqref{eq:convergence_scaling} also shows that if $\left | \alpha_1 \right |$ and $\left | \alpha_2 \right |$ are close to unity; i.e., if the estimates are close to the true value, the rate of convergence becomes quite slow. Slow convergence close to the true solution has indeed been observed in our simulations.

Another interesting insight from this analysis lies in the distribution of consecutive estimates around the true value. Each full iteration involves a gain estimate that scales the true gain vector with $\left | \alpha \right |$ and a gain estimate scaling $1/ \left | \alpha \right |$ with $\left | \alpha \right |$ moving closer to unity in each full iteration. The algorithm thus generates two sequences of points that converge to the true value, one from above and one from below.

%
%
\subsection{Convergence in general}

To assess the convergence in the general case of calibration on an arbitrary scene, we first note that the condition for convergence in \eqref{eq:convergence_req2} is tightest if either one of the terms on the righthand side equals zero. We therefore analyze those two extreme cases. In the first case, $\bepsilon = \bzero$, we have $\bbeta_1 = \widetilde{\bbeta}_1 = \bone$. This leads to the condition expressed in \eqref{eq:convergence_req_pointsrc}, for which convergence was discussed in the previous subsection.

In the second case $\alpha_1 = \alpha_2 = \alpha$, making the first term of the righthand side of \eqref{eq:convergence_req2} zero, such that the condition for convergence holds if
\begin{equation}
\frac{1}{4} \left | \alpha \left ( \widetilde{\beta}_{1, p_\mathrm{max}} - 1 \right ) + \frac{1}{\overline{\alpha}} \left ( \beta_{1,p_\mathrm{max}}  - 1 \right ) \right |^2 \leq \left | \alpha \right |^2 \frac {\left \Arrowvert \bepsilon \right \Arrowvert^2}{\left \Arrowvert \bg \right \Arrowvert^2}. \label{eq:convergence_general_req}
\end{equation}
Since $\beta_{p_\mathrm{max}}$ is based on a weighted inner product instead of the usual Euclidean inner product (as in the single point source case), we cannot show that \eqref{eq:convergence_general_req} holds in general. The significance of this is that it is possible to construct cases for which the algorithm fails. For example, when $\bM_{:,p} = \bzero$ for some value of $p$ (this holds for all $p$ in the case of an empty scene) or if, for specific weights $\bw_p$, $\left \Arrowvert \bg + \bepsilon \right \Arrowvert_{\bw_p} \rightarrow 0$, $\beta_{p_\mathrm{max}}$ becomes very large.

In summary, we conclude that, to ensure convergence, the following conditions should be met:
\begin{enumerate}
\item The inner product between the initial estimate and the true value should be nonzero; i.e., the initial estimate should not lie in the space orthogonal to the true value or be close to the zero vector.
\item The observed scene should be such that $\gamma$ in $\left \langle \bepsilon, \bg \right \rangle_{\bw_p} = \gamma \left \Arrowvert \bepsilon \right \Arrowvert_{\bw_p} \left \Arrowvert \bg \right \Arrowvert_{\bw_p}$ is small for all values of $p$. This ensures that $\beta_{p_\mathrm{max}}$ remains small. Physically, this means that the observed scene should be suitable for a calibration measurement. For example, it is not possible to do gain amplitude and phase calibration on a homogeneously filled or empty scene.
\end{enumerate}
The first criterion can usually be ensured by available knowledge of the measurement system, either from precalibration or design. The second criterion is a requirement for any calibration measurement. We therefore conclude that the algorithm will work in most practical situations. This conclusion was confirmed by extensive testing in simulation, of which some examples will be presented in the next section.

%
%
\section{Simulations}
\label{sec:performance}

In this section we show how the algorithm performs in terms of numerical, computational, and statistical performance.

%
%
\subsection{Numerical performance and practical convergence}
\label{ssec:num_perf}

We tested StEFCal with an extensive number of cases that cover a wide range of simulated sky models, with varying numbers and locations of receivers and levels of corruption and noise. These simulations all supported the conclusions drawn from the specific cases reported here.

\begin{figure}
\centering
\includegraphics[width=0.85\columnwidth]{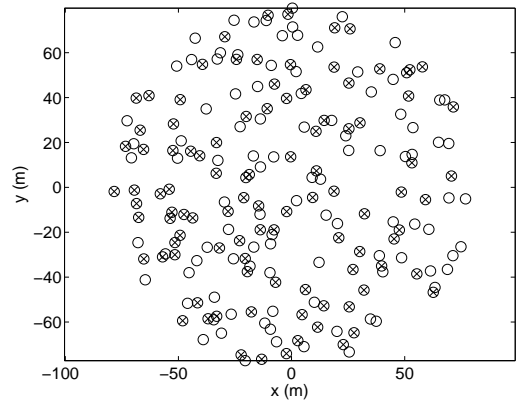}
\caption{Positions of the first $200$ (circles) and $100$ (crosses) antennas in the random configuration of $4,000$ antennas generated for the simulations. \label{fig:antennas}}
\end{figure}

The results shown here, unless otherwise indicated, employ a simulated sky consisting of $1,000$ point sources with powers exponentially distributed between $10^0$ and $10^{-4}$ Jy (1 Jy equals $10^{-26}$ W/m$^2$/Hz) randomly positioned in the sky. The array configuration consists of up to $4,000$ antennas randomly distributed over a circular range with a diameter of 160 m and minimum separation of 1.5 meters, corresponding to a Nyquist frequency of 100 MHz. Where a smaller number of antennas was required, the upper left portion of the visibility matrix (associated with the first $P$ antennas) was used, as illustrated by the antenna layouts in Fig.\ \ref{fig:antennas}. For convenience of presentation, uncorrelated receiver noise was not included in the example illustrated in this section. The effect of noise is studied in more detail in Sec.\ \ref{sec:sims}.

The model visibility matrix was built for two cases:
\begin{description}
\item[Case 1:] Incomplete sky model: only the 18 brightest sources (all sources brighter than $1\%$ of the brightest source) were included in the model visibilities.
\item[Case 2:] Complete sky model: all $1,000$ sources were included.
\end{description}

Case 1 represents a typical situation in practical radio astronomical calibration scenarios. For both cases, convergence to machine accuracy ($10^{-15}$) and to a much looser tolerance ($10^{-5}$) was studied and compared to give an indication of the convergence requirements in realistic cases.

\begin{figure}
\centering
\includegraphics[width=0.85\columnwidth]{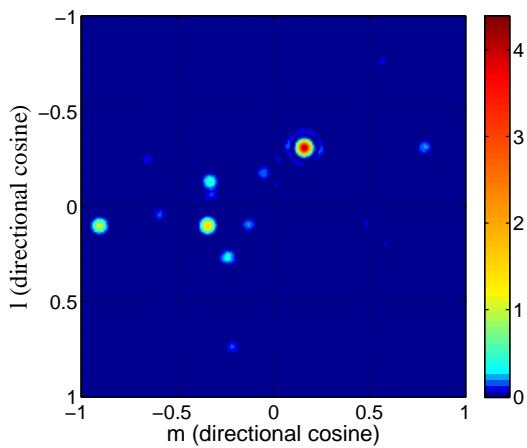}
\caption{Scene as observed by a perfectly calibrated array. The sky model contains $1,000$ sources but only the brightest are visible with this color scale. The weakest sources are even drowned in the sidelobe response of the brightest sources. \label{fig:exact_sky}}
\end{figure}
\begin{figure}
\centering
\includegraphics[width=0.85\columnwidth]{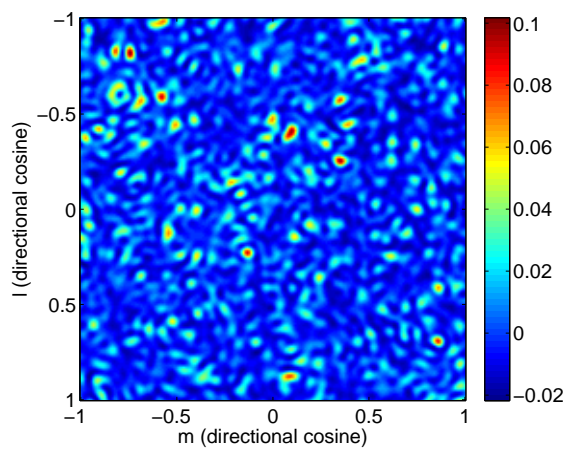}
\caption{Observed sky: sky image as seen by the instrument prior to calibration \label{fig:observed_sky}}
\end{figure}
\begin{figure}
\centering
\includegraphics[width=0.85\columnwidth]{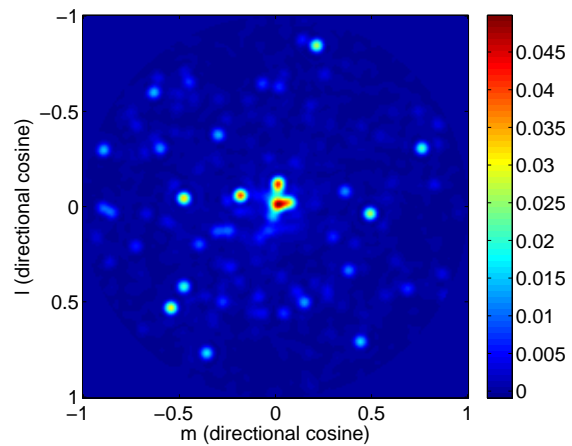}
\caption{Difference between the image after calibration and the model sky containing only the 18 brightest sources.
\label{fig:diff_calibrated_model_sky}}
\end{figure}

\begin{figure}
\centering
\includegraphics[width=0.85\columnwidth]{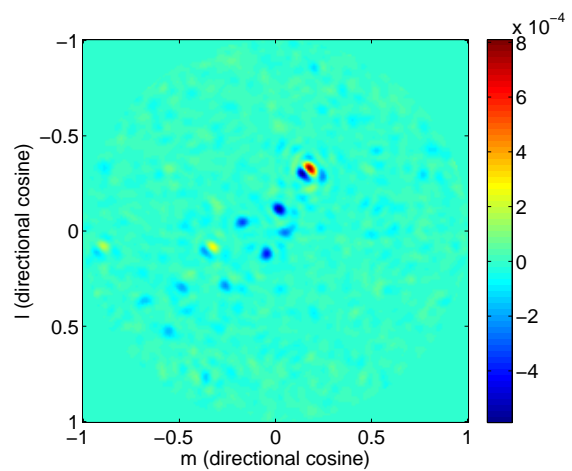}
\caption{Difference between the image after calibration on only the 18 brightest sources and the exact sky. \label{fig:diff_calibrated_exact_sky}}
\end{figure}

Figure \ref{fig:exact_sky} shows the "exact" sky, i.e. the sky image as viewed by the perfectly calibrated array. "Corrupted" skies were obtained by perturbing the antenna-based gains with amplitudes randomly distributed between 0.5 and 1.5 with unit mean and phases randomly distributed between 0 and $2\pi$ radians. The results for Case 1 are shown in Figures \ref{fig:observed_sky} through \ref{fig:diff_calibrated_exact_sky}, using \mbox{$P = 500$} antennas and at a frequency of about 35.5 MHz. Figure \ref{fig:observed_sky} shows the sky as imaged by the instrument prior to calibration. We ignored the autocorrelations of the measured visibility matrix and the matrix of model visibilities. As discussed in Sec.~\ref{sec:discussion}, it is straightforward to use StEFCal in scenarios with more entries of the covariance matrix set to zero to flag specific data values. This was studied, but not reported here for reasons of brevity, and supports the conclusions drawn.

The images obtained after calibration are indistinguishable from Fig.\ \ref{fig:exact_sky} at the resolution of the array, so they are not shown here. The difference between the image after calibration and the incomplete sky model (which includes just 18 sources for Case 1) is more interesting and is shown in Fig.\ \ref{fig:diff_calibrated_model_sky}. This clearly shows that the next brightest sources can be identified correctly after calibration and subtraction of the 18 brightest sources.
Figure \ref{fig:diff_calibrated_exact_sky} shows the difference between the image after calibration and the exact sky showing that the errors due to ignoring the weak sources in the model are two orders of magnitude smaller than the next brightest sources. This shows that StEFCal can be used as an algorithmic component in self-calibration procedures based on iterative source model refinement.

\begin{table}
\centering
\caption{Maximum difference between exact sky and image after calibration to different tolerances. \label{tab:cal_diff}}
\begin{tabular}{|l|r|}
\hline
\hline
max (sky\textsubscript{exact} - sky\textsubscript{$10^{-5}$}) & $3.2\cdot10^{-8}$\\
max (sky\textsubscript{exact} - sky\textsubscript{$10^{-15}$}) & $1.3\cdot10^{-15}$\\
max (sky\textsubscript{$10^{-5}$} - sky\textsubscript{$10^{-15}$}) & $3.2\cdot10^{-8}$\\
\hline
\hline
\end{tabular}
\end{table}

We also carried out simulations with different settings for the tolerance for convergence. The results for Case 2 (complete sky model) are reported here. Table \ref{tab:cal_diff} summarizes the results for
tolerances of $10^{-5}$ and $10^{-15}$. In the table, we show the maximum difference between the absolute value of the image pixels between the exact sky and the image after calibration for both convergence tolerances. The maximum difference we found is on the order of $10^{-8}$, which is well below the noise level. This led us to conclude that effective convergence can be achieved with limited effort and reasonably "loose" convergence requirements. The table also reports the difference between two images obtained after calibration for the two tolerances.

As already mentioned, we have run a long series of tests with the number of antennas ranging from $20$ to $4,000$, a variety of source models, with and without adding antenna-based noise. All cases showed similar outcomes to those reported here. This supports the suitability of StEFCal for practical use in terms of numerical performance and speed of convergence. The number of iterations required appears to depend only very weakly on the number of receivers. Typically, in the cases reported here, irrespective of the number of antennas, $10^{-5}$ or better convergence could be achieved in 20 iterations or less and $10^{-15}$ convergence in 40 iterations or less. Benchmarks are reported in Sec. \ref{subsec:perf}.

%
%
\subsection{Computational performance}
\label{subsec:perf}

The incomplete source model introduced in Sec. \ref{ssec:num_perf} (Case 1) was used in the computational benchmarks reported here, although far more extensive tests were carried out. In all cases, StEFCal has shown very good performance characteristics despite its iterative nature. This requires a refresh of the data for each iteration for problems too large to fit in cache, as is the case for any other iterative approach.

A number of factors underpin StEFCal's performance:
\begin{itemize}
\item All computations are carried out through vector operations.
\item Data are accessed by unit-stride, contiguous memory patterns.  This ensures maximum utilization of data loaded, ensuring full use of cache lines, as well as emphasizing the role of prefetching.
\item StEFCal requires only a modest memory footprint (as discussed below). 
\end{itemize} 

We coded the algorithm in Fortran 90, compiled using the Intel MKL Fortran compiler version 11.00. We used either Intel MKL BLAS or handwritten code. The difference between these two versions was 3\% at most, so we only report on the results using the Intel MKL library. We used a desktop system with an Intel dual core Core 2 running at 3.0GHz, single threaded (only one core active), with single-threaded MKL BLAS.

We would like to mention that multithreaded parallelism over frequency channels and/or time slices have also been developed and tested to very good effect.
\begin{table}
\centering
\caption{Algorithm measured performance.\label{tab:benchmarks}}
\begin{tabular}{r|r|r|r|r|r}
\hline
\hline
 & \multicolumn{2} {|c} {$N_{it}$ fixed} & \multicolumn{3}{|c}{Tolerance 10\textsuperscript{-5}} \\
 \hline
$P$ & $N_{it}$ & time & $N_{it}$ & time & convergence \\
\hline
50   & 40 & 0.0006 & 12 & 0.0002 & 3.5E-06 \\
100  & 40 & 0.002 & 14 & 0.001 & 4.9E-06 \\
200	& 40	 & 0.008 & 16 & 0.003 & 1.9E-06 \\
300 & 40	 & 0.017 & 16 & 0.007 & 3.6E-07 \\
400  & 40 & 0.036 & 16 & 0.014 & 2.6E-06 \\
500  & 40 & 0.062 & 18 & 0.028 & 3.0E-08 \\
600  & 40 & 0.093 & 18 & 0.042 & 1.7E-08 \\
800  & 40 & 0.220 & 18 & 0.099 & 5.5E-07 \\
1000 & 40 & 0.258 & 18 & 0.116 & 3.8E-08 \\
1500 & 40 & 0.579 & 18 & 0.260 & 1.9E-07 \\
2000 & 40 & 1.029 & 20 & 0.515 & 4.5E-09 \\
3000 & 40 & 2.321 & 20 & 1.169 & 1.6E-08 \\
4000 & 40 & 4.108 & 20 & 2.059 & 1.6E-07 \\
\hline
\hline
\end{tabular}
\tablefoot{
In the table, $N_{it}$ is the number of iterations and all times are in seconds.
}
\end{table}
\begin{figure}
\centering
\includegraphics[width=0.85\columnwidth]{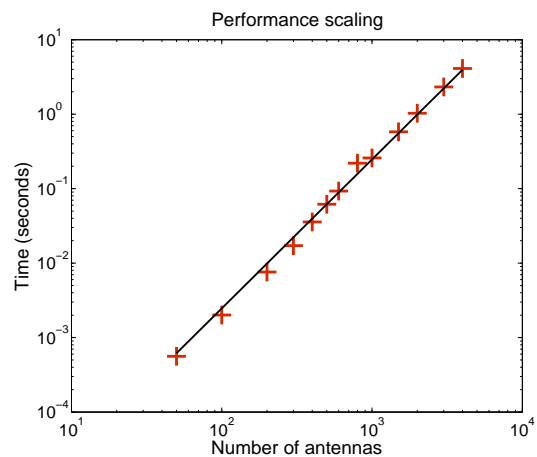}
\caption{Algorithm scaling on a logarithmic scale. The red crosses denote the measured computing times; the black line shows the expected computing times for quadratic scaling, normalized to $P = 500$:
$T_P = (\frac{P}{500})^2 \cdot T_{500}$.}
\label{fig:scaling}
\end{figure}
Table \ref{tab:benchmarks} and Fig.\ \ref{fig:scaling} report the performance obtained for arrays with $50$ to $4,000$ antennas. The figure compares the measured computing times against those expected for
perfect quadratic scaling with the number of antennas, normalized to $P = 500$. Both the table and figure highlight the efficiency of the algorithms, as well as its quadratic scalability over the number of elements.

One full iteration requires the element-wise product of two complex vectors and two complex dot products per gain parameter. In terms of {\em real valued} floating point operations, this is equivalent to $24 P^2$ flop, where each complex multiply-add requires 8 flop. One dot product corresponds to computing the square of the 2-norm (or F-norm)  and the cost of that dot product can be halved, leading to a grand total per iteration of $20 P^2$ flop. The algorithm does indeed require $\mathcal{O} (P^2) $ operations as in our initial claim.

The system used for benchmarking had a peak speed of 12 Gflops, and the peak speed observed for MKL DGEMM was $\sim$11 Gflops. StEFCal showed performance figures of over 3 Gflops. Given the nature of the computation, both the speed and the scalability observed are very good (over 25\% peak speed). Tests have shown that even better performance can be obtained on more modern CPUs.

The memory footprint of StEFCAL is modest:
\begin{itemize}
\item The measured visibilities and the model visibilities are complex valued $P \times P$ matrices, requiring $16 P^2$ bytes per matrix for storage in double-precision floating-point format. Thanks to their Hermiticity, one could store these matrices in compressed triangular storage format. However, this would require accessing their elements with non-unit variable strides, thus considerably lowering computational performance: given the
memory available on current systems, performance issues are overriding.
\item One complex valued vector of length $P$ is returned as output.
\item One internally allocated complex valued vector of length $P$ is used.
\item Depending on code internals, other complex vectors of length $P$ may be required.
\end{itemize}

The total amount of input and output data ($32 P^2 + 16 P$ bytes assuming  double precision floating point format) results in the low computational intensity of $24 P^2 N_{it} / \left ( 32 P^2 + 16 P \right )$ flop per byte or approximately $3 N_{it} / 4$ flop per byte for high values of $P$. Thus StEFCal is memory bound, as observed in practice. As already mentioned, this is greatly ameliorated by the memory access pattern. While traditional $\mathcal{O}(P^3)$ methods may increase the number of operations per memory transfer, they also increase the number of operations by the same amount, thus resulting in much lower performance overall.

%
%
\subsection{Statistical performance}
\label{sec:sims}

We performed a series of Monte Carlo simulations to assess the statistical performance and robustness of the proposed algorithm. We have defined three scenarios:
\begin{enumerate}
\item Calibration on two weak (S/N of -10 dB per time sample) point sources.
\item Calibration on a realistic sky model.
\item Calibration on two strong (S/N of 20 dB per time sample) point sources.
\end{enumerate}
By choosing these scenarios, we try to explore the sensitivity of StEFCal to violation of the assumption that $\bR \approx \sigma_n \bI$, which was used to derive the algorithm from the weighted least squares cost function and hence get a feel for the range of applicability of the algorithm. The antenna configuration used in these simulations is the 200-element configuration shown in Fig.\ \ref{fig:antennas}.

Since we can only estimate the phase difference between the antennas, we assume that the first antenna will be used as phase reference. We can therefore define the $\left ( 3P - 1 \right ) \times 1$ vector of free parameters $\btheta = \left [ \gamma_1, \cdots, \gamma_P, \varphi_2, \cdots, \varphi_P, \sigma_{n,1}, \cdots, \sigma_{n,P} \right ]^T$, where $\gamma_p$, $\varphi_p$, and $\sigma_{n,p}$ are the gain amplitude, gain phase, and the noise power of the $p$th element, respectively. Expressions for the Cramer-Rao bound (CRB), the minimum achievable variance for an unbiased estimator \citep{Kay1993, Moon2000}, for this scenario is derived by \citet{Wijnholds2009TSP}.

\subsubsection{Calibration on two weak point sources}

In this scenario, two sources with source power $\sigma_q = 1$ for $q = 1, 2$ were located at $\left ( l_1, m_1 \right ) = \left ( 0, 0 \right )$ (field center or zenith) and $\left ( l_2, m_2 \right ) = \left ( 0.4, 0.3 \right )$, where $l$ and $m$ are direction cosines. We set the measurement frequency to 60 MHz ($\lambda = 5.0$ m) and defined $\sigma_n = 10$ for all antennas, such that both sources have an S/N of -10 dB per time sample. This ensures that the assumption $\bR \approx \bsigma_n \bI$, which we used to derive the algorithm from the weighted LS cost function, holds. In this Monte Carlo simulation, we calibrated the data using StEFCal, as well as the multisource calibration algorithm proposed in \citet{Wijnholds2009TSP}, to compare the two approaches in terms of statistical and computational efficiency. Simulations were done for $K = \left \{ 10^3, 3 \cdot 10^3, 10^4, 3 \cdot 10^4, 10^5, 3 \cdot 10^5, 10^6 \right \}$ time samples and each simulation was repeated 100 times. For Nyquist-sampled time series, the number of samples is equal to the product of bandwidth and integration time, i.e., $K = B\tau$. The chosen range of values for $K$ thus covers the most commonly used range of values for bandwidth and integration time in radio astronomical calibration problems with high spectral and temporal resolution.

The biases found in the simulations are considerably smaller than the standard deviation for this scenario based on the CRB. This indicates that our algorithm is unbiased, hence that a comparison with the CRB is meaningful to assess the statistical performance of the algorithm.
\begin{figure}
\centering
\includegraphics[width=0.85\columnwidth]{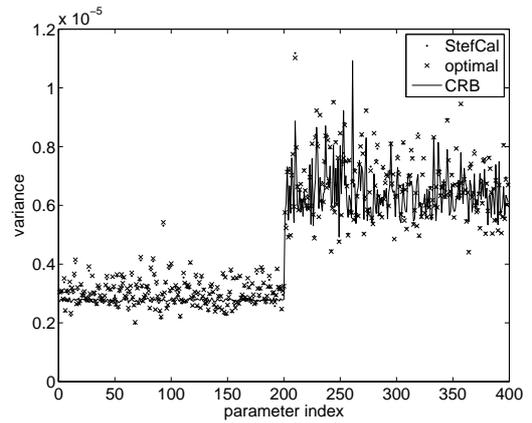}
\caption{Variance on the estimated gain amplitudes (indices 1 through 200, dimensionless) and phases (indices 201 through 399, in units of rad$^2$) parameters for calibration on two point sources with a S/N of -10 dB. The solid line indicates the CRB. \label{fig:var_allparams}}
\end{figure}
\begin{figure}
\centering
\includegraphics[width=0.85\columnwidth]{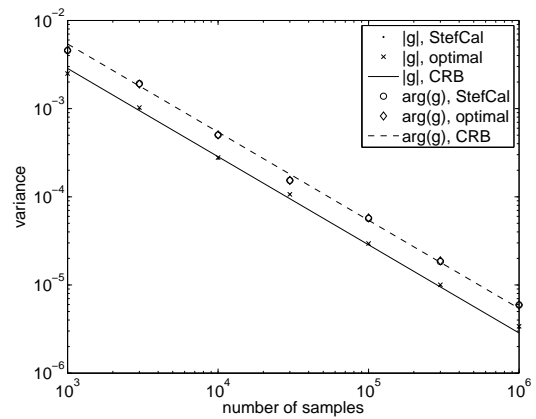}
\caption{Variance (dimensionless for amplitude parameters, in units of rad$^2$ for phases) of a representative complex valued gain estimate ($p = 10$) as function of the number of time samples. The lines mark the CRB for the two parameters involved. \label{fig:var_vs_K}}
\end{figure}
Figure \ref{fig:var_allparams} shows the variance of the estimated gain amplitude and phase parameters for $K = 10^6$, clearly showing that both algorithms achieve the CRB for large $K$. Figure \ref{fig:var_vs_K} shows the variance for a representative complex valued gain estimate for all simulated values of the number of samples $K$. The result indicates that the CRB is already achieved for very low values of $K$. Based on the theory of random matrices, matrix-wise convergence of the covariance matrix estimate starts when the number of samples is about ten times bigger than the number of elements \citep{Couillet2011ICASSP}, which, in the case of a 200-element array, would be at $K \approx 2 \cdot 10^3$. The proposed algorithm does not rely on mathematical operations that depend on matrix-wise convergence to work properly, and this may provide an intuitive explanation for this attractive feature of StEFCal.

The simulation results indicate that StEFCal achieves statistically optimal performance when $\bR \approx \sigma_n \bI$ and thus has statistical performance similar to statistically efficient methods, such as the algorithm described by \citet{Wijnholds2009TSP} or optimization of the cost function using the Levenberg-Marquardt solver. However, the proposed algorithm has only $\mathcal{O}(P^2)$ complexity instead of the $\mathcal{O}(P^3)$ complexity of many commonly used methods. This should give a significant reduction in computational cost of calibration, especially for large arrays. The Monte Carlo simulations described here were done in Matlab on a single core of a 2.66 GHz Intel Core i7 CPU. Gain calibration for a single realization took, on average, 2.24 s when using the method described by \citet{Wijnholds2009TSP} while taking only 0.12 s when using StEFCal.

\subsubsection{Calibration on a realistic scene}

In many array applications, the scene on which the array needs to be calibrated in the field is considerably more complicated than one or just a few point sources. To see how the algorithm performs in a more realistic scenario, we used the scene shown in Fig.\ \ref{fig:exact_sky} for calibration. We are still in the low-S/N regime, with the noise power in each antenna being about ten times the total power in the scene and the strongest sources having an S/N per sample of -13.3 dB, so $\bR \approx \sigma_n \bI$ still holds.
\begin{figure}
\centering
\includegraphics[width=0.85\columnwidth]{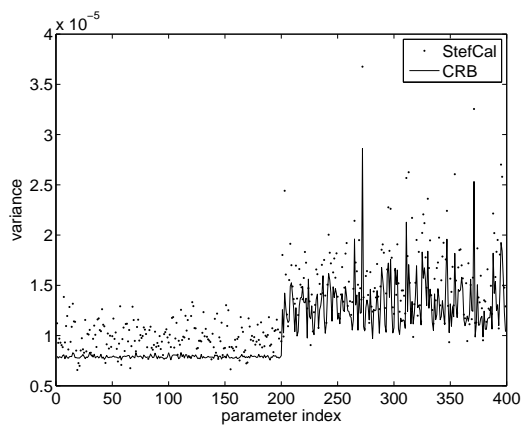}
\caption{Variance on the estimated gain amplitudes (indices 1 through 200, dimensionless) and phases (indices 201 through 399, in units of rad$^2$) parameters for calibration on the scene shown in Fig.\ \ref{fig:exact_sky} with noise power equal to the integrated power of all sources. \label{fig:var_allparams2}}
\end{figure}

\begin{figure}
\centering
\includegraphics[width=0.85\columnwidth]{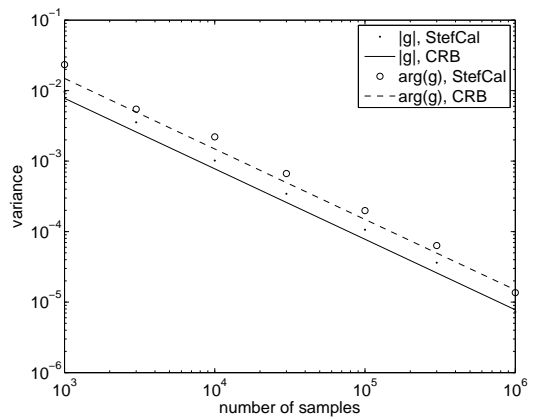}
\caption{Variance (dimensionless for gain amplitude, in units of rad$^2$ for gain phase) of a representative complex valued gain estimate ($p = 20$) as function of the number of time samples. The lines mark the CRB of the two corresponding parameters. \label{fig:var_vs_N2}}
\end{figure}

We set up our Monte Carlo simulations in the same way as for the first scenario. After checking that the algorithm produced unbiased results, we compared the variance on the estimated parameters with the CRB. The results are shown in Figs.\ \ref{fig:var_allparams2} and \ref{fig:var_vs_N2}. They indicate that the performance of StEFCal is still very close to statistically optimal.

\subsubsection{Calibration on two strong point sources}

For our last scenario, we defined a simulation with two point sources located at $(l_1, m_1) = (0, 0)$ and $(l_2, m_2) = (0.4, 0.3)$ with an S/N of 20 dB per time sample. This is a scenario that clearly violates the assumption that $\bR \approx \bsigma_n \bI$. We performed the Monte Carlo simulation in the same way as in the previous cases.

\begin{figure}
\centering
\includegraphics[width=0.85\columnwidth]{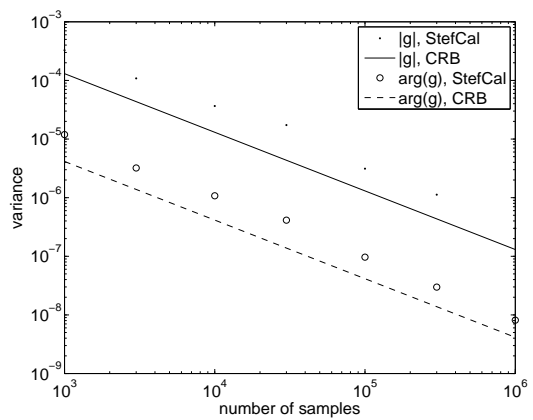}
\caption{Variance (dimensionless for gain amplitude, in units of rad$^2$ for gain phase) of a representative complex valued gain estimate ($p = 5$) as function of the number for time samples. The lines mark the CRB of the two corresponding parameters. \label{fig:var_vs_N4}}
\end{figure}

Figure \ref{fig:var_vs_N4} shows the variance of the gain and phase associated with a representative element as function of the number of samples. The gain estimates, while not as close to the CRB as in the previous cases, are still quite close to the bound. The average gain amplitude error is still only $55\%$ higher than the CRB, while the average phase error is only $26\% higher$ than the CRB. This is acceptable given the high accuracy achieved in such a high-S/N regime. We conclude that StEFCal provides a performance that is close to optimal, even in scenarios designed to break the underlying assumptions made to use the LS rather than the WLS cost function. This shows that the algorithm is fairly robust in terms of its statistical performance and will provide statistically efficient estimates in scenarios typical of radio astronomy.

%
%
\section{Extension to full polarization calibration}
\label{sec:fullpol}

It is straightforward to apply the ADI approach to the full polarization case as demonstrated by \citet{Hamaker2000AandA} and \citet{Mitchell2008JSTSP}. Initial results for StEFCal have been presented by \citet{Salvini2014CALIM} and \citet{Salvini2014URSI} and confirm the validity of StEFCal for full polarization calibration, still retaining $\mathcal{O}(P^2)$ computational complexity. In this section, we sketch the StEFCal algorithm for the full polarization case. A full analysis will be provided in a future paper.

The mathematical problem is structured in terms of matrices whose elements are two-by-two complex blocks, rather than by individual complex values. In particular, the gain matrix $\bG$ is a block diagonal matrix whose $2 \times 2$ blocks on the main diagonal describe the response of the two feeds of each receiver:
\begin{equation}
\bG_p  =
   \left[
      \begin{array}{ll}
         \bG_{2p-1,2p-1} & \bG_{2p-1, 2p} \\
         \bG_{2p,2p-1} & \bG_{2p, 2p}
      \end{array}
   \right] .
\end{equation}

Taking this structure into account, the full polarization calibration problem can still be formulated as
\begin{equation}
\widehat{\bG} = \underset{\bG}{\argmin} \left \| \widehat{\bR} - \bG \bM \bG^H \right \|_F^2 .
\end{equation}
It naturally follows that the basic step of full polarization StEFCal consists of solving $P$ $2 \times 2$ linear least squares problems for each iteration, within the same StEFCal iteration framework as for the scalar case described in this paper.

In general, the simple StEFCal algorithm, which has proved very successful for the scalar case, exhibits slow or difficult convergence. As shown by \citet{Salvini2014URSI}, this can be corrected by employing a multistep approach, whereby the two previous solutions at the even steps are also included in the averaging process. Moreover, some heuristics can be employed to regularize the convergence rate.

Performance again proved very good, since the same considerations as for the scalar algorithm apply. Since the density of operations increases by a factor two per data item, a marginally better speed has been obtained, in terms of Gflop per second.
An example of performance results is shown in Table \ref{tab:polbenchmarks}. This involved a realistic scenario of full polarization calibration of the proposed SKA Low Frequency Aperture Array (LFAA) station, comprising of 256 antennas ($512$ dipoles) for $1024$ frequencies. The code was parallelized over frequencies using OpenMP, whereby each core grabs the first available frequency still to be calibrated (dynamic load balancing). All computations were carried out using single precision to a tolerance of 10$^{-5}$, delivering better than 1\% accuracy in the complex gains, as required. Performance figures are compared to the performance of MKL CGEMM (complex matrix - complex matrix product), which virtually runs at peak speed and gives a good indication of maximum speeds achievable.

\begin{table}
\centering
\caption{Full polarization StEFCal performance (see text). Times are in seconds. \label{tab:polbenchmarks}}
\begin{tabular}{r|r|r|r}
\hline
\hline
N. cores & time & GFlops/sec & \% CGEMM \\
\hline
1 & 16.63 & 17.9 & 41.8\\
2 & 8.34 & 35.8 & 41.9\\
3 & 5.57 & 53.5 & 42.2\\
4 & 4.19 & 71.2 & 41.8\\
8 & 2.90 & 102.9 & 42.0\\
10 & 2.35 & 127.0 & 42.2\\
12 & 1.92 & 155.3 & 41.3\\
16 & 1.72 & 173.3 & 38.3\\
\hline
\hline
\end{tabular}
\tablefoot{
In the table, all times are in seconds.
}
\end{table}

\begin{figure}
\centering
\includegraphics[width=0.85\columnwidth]{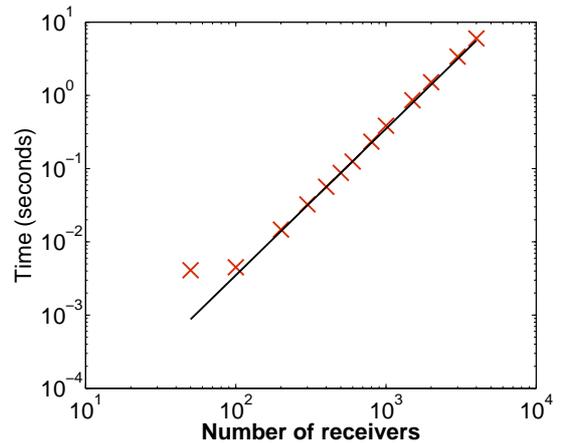}
\caption{Full polarization algorithm scaling on a logarithmic scale. The red crosses denote the measured computing times; the black line shows the expected computing times for quadratic scaling, normalized to $P = 500$: $T_P = (\frac{P}{500})^2 \cdot T_{500}$.}
\label{fig:polscaling}
\end{figure}

Figure \ref{fig:polscaling} shows that scalability with problem size has very similar characteristics to those for the scalar problem. In this simulation, we fixed the number of iterations to $100$ and compared the measured data against exact $P^2$ scalability, normalized to $P = 500$. It should be noted that the number of operations per iteration now reads as $48 P^2$. We also like to point out that the convergence rate, i.e. the number of iterations required for a given accuracy, exhibits the same very weak dependence on problem size, i.e. the number of receivers, as in the case of scalar StEFCal.

%
%
\section{Applications}
\label{sec:practice}

\begin{figure}
\centering
\includegraphics[width=0.94\columnwidth]{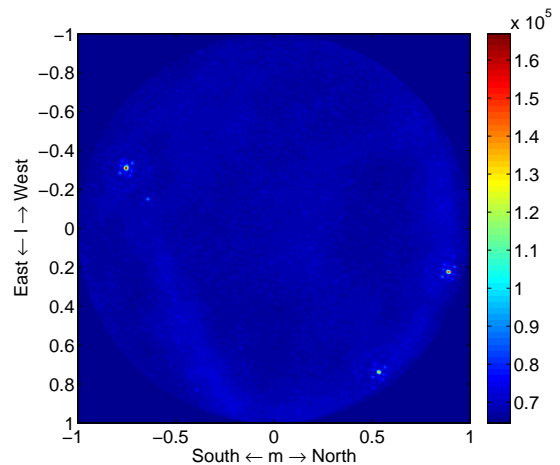}
\caption{Calibrated all-sky image at 59.67 MHz made with the 288-antenna AARTFAAC system. \label{fig:AARTFAAC_map}}
\end{figure}

Figure \ref{fig:AARTFAAC_map} shows a calibrated all-sky map produced by the Amsterdam-ASTRON Radio Transient Facility and Analysis Centre (AARTFAAC, www.aartfaac.org, \citet{Prasad2013RSTA}). This system is a transient monitoring facility installed on the six innermost stations of the LOFAR, which this facility combines as a single station with 288 antennas spread over an area with a diameter of about 350 m. In this section, we describe the integration of StEFCal in the AARTFAAC calibration pipeline to demonstrate the ease of integration of the proposed algorithm in an existing pipeline and the computational benefits.

\begin{table}
\centering
\caption{Processing time in seconds per main iteration of array calibration using the original WALS as described by \citet{Wijnholds2010PhD} and WALS with StEFCal. \label{tab:compute_load}}
\begin{tabular}{l|r|r}
\hline
\hline
parameter & original & with StEFCal\\
\hline
gain & 6.528 & 0.184\\
source power & 0.023 & 0.030\\
noise covariance & 0.003 & 0.003\\
\hline
total & 6.554 & 0.217\\
\hline
\hline
\end{tabular}
\end{table}

The calibration of the AARTFAAC system involves estimating the direction-independent complex valued gains of the receiving elements, the apparent source powers of the four bright point sources seen in Fig.\ \ref{fig:AARTFAAC_map}, and a non-diagonal noise covariance matrix to model the diffuse emission seen in the image and the system noise. The non-diagonal noise covariance matrix was modeled with one complex valued parameter for every entry associated with a pair of antennas that were less than ten wavelengths apart and a real valued parameter for every entry on the main diagonal. This calibration challenge was solved using the weighted alternating least squares (WALS) algorithm described by \citet{Wijnholds2010PhD}. In each main iteration of the WALS method, the direction-independent gains are first estimated assuming that the other parameters are known, then the source powers are updated, and finally the noise covariance matrix is updated. To calibrate the AARTFAAC data set used to produce Fig.\ \ref{fig:AARTFAAC_map}, six main iterations were required. The middle column of Table \ref{tab:compute_load} shows the average time estimation of each group of parameters took in Matlab on an Intel Core i7 CPU on a machine with 4 GB RAM.

StEFCal was easily integrated in the WALS algorithm by simply replacing the gain estimation step (which used an algorithm of $\mathcal{O} \left ( P^3 \right )$ complexity) with the StEFCal algorithm.  StEFCal was configured to iterate to convergence in each main loop of the WALS method. Table \ref{tab:compute_load} reports the computational times, showing that StEFCal resulted in an increase in performance of over a factor $30$ when "cold starting", i.e. without any prior information for any timeslice. As gains are expected to vary smoothly over time, a further eight-fold increase in performance was obtained by using the results from the previous timeslice as initial guess in the calibration of each snapshot (a factor 250+ overall). This underscores the capability of StEFCal to make good use of initial gain estimates.

The full polarization version of StEFCal is currently employed in MEqTrees \citep{Noordam2010AandA}. It is being implemented in the standard LOFAR pre-processing pipeline and studied for the SKA. As an example, the LOFAR central processor requires a number of steps including direction-independent gain calibration (station-based) to provide initial corrections for clock drift and propagation effects. The latter are mainly caused by the ionosphere and may require full polarization corrections on baselines to stations outside the core area. Recently, \citet{Dijkema2014CalIm} has implemented the basic version of full polarization StEFCal for the standard processing pipeline of LOFAR. This implementation was used to run the same pipeline on several data sets from actual LOFAR observations twice, once with the standard Levenberg-Marquardt (LM) solver and once with the LM solver replaced by StEFCal. In all cases, the results obtained were practically identical, but the pipeline with StEFCal was typically a factor 10 to 20 faster than the pipeline running the LM solver. Based on the material presented in this paper, we expect that we can improve performance significantly by optimizing the implementation of StEFCal used.

\section{Discussion and future work}
\label{sec:discussion}

\subsection{Other variants of StEFCal}

We also studied a variant of StEFCal with relaxation, in which the complex gains are used as soon as they become available, rather than using the full set of complex gains from the previous iteration; i.e. the gain vector gets updated while looping over all receivers and is then applied immediately. This variant is listed in Algorithm \ref{alg:stefcal2}. In general, this variant needs fewer iterations. However, the receiver loop (the $p$-loop) for each iteration has parallel dependencies, which makes this variant much less portable to multicore and many-core platforms, such as GPUs, although it could be valuable for more traditional CPUs.

\begin {algorithm}
\caption{Algorithm StEFCal2}
\begin{algorithmic}
\State Initiate $\bG^{[0]}$;  $\bG^{[0]} = \bI$ is adequate in most cases
\For { $i  =  1, 2, \cdots , i_\mathrm{max}$} 
\State $\bG^{[i]} = \bG^{[i-1]}$
\For { $p  =  1, 2, \cdots , P $} 
\State $\bz \gets \bG^{[i]}  \cdot \bM_{:, p}  \equiv \bg^{[i]} \odot \bM_{ :, p }$
\State $g_p^{[i]} \gets (\bRh_{:, p}^H \cdot \bz ) / (\bz^H \cdot \bz)$
 \EndFor
 \If {$\| \bg^{[i]} - \bg^{[i-1]} \|_F / \| \bg^{[i]} \|_F \leq \tau$} 
\State Convergence reached\EndIf
\EndFor
\end {algorithmic}
\label{alg:stefcal2}
\end {algorithm}

\begin{figure}
\centering
\includegraphics[width=0.85\columnwidth]{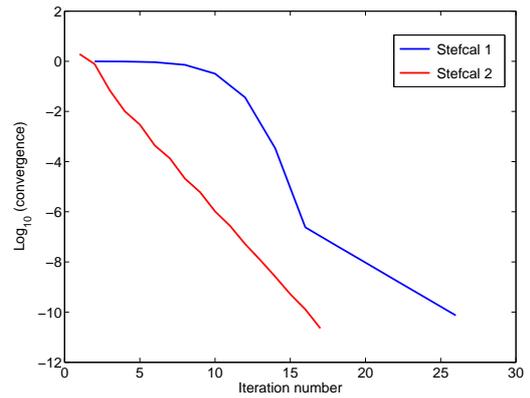}
\caption{Comparing algorithm performance for P = 500. \label{fig:stefcal_comparison}}
\end{figure}

Numerical performance appears very close to the standard StEFCal, shown as Algorithm \ref{alg:stefcal}, but we have not attempted to obtain a formal proof of convergence. Figure \ref{fig:stefcal_comparison} shows the faster convergence of Algorithm \ref{alg:stefcal2}, in particular at the beginning of the iteration.

Another variant of Algorithm \ref{alg:stefcal2}, whose details are not reported here, aims to decrease the parallel dependencies by block-wise updates of the estimated gain vector (thus the latest values of the previous block are used, while the old values of the current block are used). As expected, performance falls in between the full relaxation and the original algorithm. It is felt by the authors that loss of parallelism is more important than a rather modest gain in performance.

The first version of StEFCal included an initial stage (again with operation count $\mathcal{O}(P^2)$), which purified the largest eigenvalues and vectors of the observed visibilities (including estimation of the autocorrelation terms), and then matched these against the corresponding ones in the model sky. This worked very well and was very fast, when the number of bright sources present in the field of view was much smaller than the array size $P$. This variant was presented at various meetings but was then dropped because of the simplicity and power of the current version of StEFCal.

\subsection{Extension to iterative reweighted least squares}

In Sec. \ref{ssec:meq}, it was pointed out that appropriate weighting of each data point may be required to reduce the effects of outliers. Typically, this is done by assigning weights to the data values based on their reliability or use a different norm, for example the 1-norm, that is less sensitive to outliers. To handle such cases, we have developed a variant of StEFCal that follows an iterative reweighed least squares (IRLS) approach \citep{Moon2000}. In an IRLS algorithm, the data values are weighted so that the 2-norm minimization becomes equivalent to minimization using another norm. In our example below, we minimize the 1-norm of the residuals. The resulting algorithm still has $\mathcal{O}(P^2)$ complexity, but the individual iterations require more operations to calculate and apply the weights.

We aim to minimize the 1-norm of the residuals, which is not the matrix 1-norm but the sum of the absolute values of all data points
\begin{equation}
\| \bDelta \|_{1} = \sum_q \sum_p | \bRh_{q, p}  -  g_q \bM_{q, p} g^*_p |,
\label{eq:1norm}
\end{equation}
where the indices $p$ and $q$ run over the receiving elements.
The kernel of the algorithm is modified by using the weights
\begin{equation}
\bW_{q, p} = \frac{1}{| \bRh_{q, p}  -  g_q \bM_{q, p} g^*_p |}
\label{eq:weights}
\end{equation}
with appropriate checks and actions for very small (or zero) entries in the denominator.

We can apply these weights in the basic StEFCal iteration by replacing \eqref{eq:adi:basic3} by
\begin{equation}
g_p^{[i]} = \frac{\widehat{\bR}_{:p}^H \cdot \left ( \bW_{:p} \odot \bZ_{:p}^{[i-1]} \right )}{\left ( \bZ_{:p}^{[i-1]} \right )^H \left ( \bW_{:p} \odot \bZ_{:p}^{[i-1]} \right )}
\end{equation}
and applying corresponding changes to Algorithm \ref{alg:stefcal}. At the end of each iteration, the appropriate weights would need to be computed using \eqref{eq:weights}.

As an example, we consider the calibration in the 2- and 1-norm of test data for 200 receivers with a tolerance of $10^{-7}$. The 2-norm calibration required 28 iterations, while the 1-norm calibration required 52 iterations. Computational costs per iteration were higher by a factor 1.5. This factor can be explained by the increased number of operations required for relatively expensive calculations like taking the absolute value of a complex number. An interesting question, which is beyond the scope of this paper, is whether it is necessary to do 1-norm optimization until convergence or whether the weights can be fixed after a limited number of iterations when the solution is sufficiently close to the optimum. Such an approach would reduce computational costs by avoiding the recalculation of weights from a certain point in the iteration process.

\subsection{Integration of StEFCal in other algorithms}

In Sec.~\ref{sec:practice}, we saw an example of how StEFCal was integrated in an existing calibration pipeline. This particular example involved a non-diagonal noise covariance matrix that was modeled by introducing an additional noise parameter for each off-diagonal entry of the noise covariance matrix that was assumed to be non-zero. In this paper, we set the diagonal entries of the array covariance matrix $\bRh$ and the model covariance matrix $\bM$ to zero. We could easily accommodate the estimation of the non-diagonal noise covariance matrix by setting not only the entries associated with the autocorrelations to zero, but also the entries associated with the non-zero off-diagonal entries of the noise covariance matrix. We can use the same procedure to account for corrupted or missing data or for when short baselines should be excluded. Of course, this should not be done unnecessarily, because exclusion of potentially useful information from the gain estimation process will degrade the gain estimation performance.

Estimation of direction-dependent gains is currently a hot topic in radio astronomy \citep{Wijnholds2010SPM}. Apart from brute force approaches using the Levenberg-Marquardt solver, two iterative approaches, the differential gains method proposed by \citet{Smirnov2011AandA} and calibration using space alternating generalized expectation maximization (SAGECal) proposed by \citet{Yatawatta2009DSP} and \citet{Kazemi2011MNRAS}, have become quite popular. Both methods iterate over the directions for which antenna-based gains need to be estimated, assuming that the other directions are already calibrated. In doing so, both methods reduce the estimation problem for each specific direction to the problem of estimating direction independent gains. Currently, the Levenberg-Marquardt solver is used for each of these subproblems, but given the nature of the problem, those estimation steps could be replaced by StEFCal, which is a solver specific to the problem. Introducing StEFCal in such an algorithm can potentially reduce their computational requirements significantly as demonstrated by \citet{Salvini2013-3GC3} and \citet{Salvini2014CALIM}.

\subsection{Summary of main results}
In this paper we have analyzed the performance of ADI methods for solving antenna-based complex valued gains with $\mathcal{O} (P^2)$ complexity. We have
\begin{itemize}
\item done a rigorous analysis of convergence showing that the algorithm converges to the optimal value except in a number of special cases unlikely to occur in any practical scenario;
\item reported on its numerical and computational performance; in particular, we highlighted its raw speed, as well as its scalability;
\item assessed the statistical performance and shown that it performs very close to the Cramer Rao bound (CRB) in most realistic scenarios;
\item commented on variations in the basic ADI method, extension to full polarization cases, and inclusion in more complex calibration scenarios.
\end{itemize}

\begin{acknowledgements}
We would like to thank Oleg Smirnov, Marzia Rivi, Ronald Nijboer, Alle-Jan van der Veen, Jaap Bregman, Johan Hamaker, and Tammo Jan Dijkema for their useful contributions during discussions, and their constructive feedback on earlier versions of this paper. The research leading to this paper has received funding from the European Commission Seventh Framework Programme (FP/2007-2013) under grant agreement No.\ 283393 (RadioNet3).
\end{acknowledgements}

\bibliography{refs}

\end{document}